\documentstyle[tighten,prl,epsf,floats,aps]{revtex} 
\def\comment#1{}

\begin{document}

\twocolumn[
\hsize\textwidth\columnwidth\hsize\csname @twocolumnfalse\endcsname

\draft

\preprint{\today}

\title{ Temperature Evolution of the Pseudogap State in the Infra-Red 
Response of Underdoped La$_{2-x}$Sr$_{x}$CuO$_{4}$} 

\author{T.~Startseva$^{(1)}$, T.~Timusk$^{(1)}$, 
A.V.~Puchkov$^{(2)}$, D.N.~Basov$^{(3)}$, H.A.~Mook$^{(4)}$, 
M.~Okuya$^{(5)}$, T.~Kimura$^{(5)}$, and  K.~Kishio$^{(5)}$} \address{$^{(1)}$ 
Department of Physics and Astronomy,  McMaster University 
 \protect \\ Hamilton, Ontario, CANADA L8S 4M1}
\address{$^{(2)}$ Department of Applied Physics, Stanford University, 
Stanford, CA 94305} \address{$^{(3)}$ Department of Physics, 
University of California San-Diego, La Jolla, CA 92093} 
\address{$^{(4)}$ Oak Ridge National Laboratory, Oak Ridge, Tennessee 
37831} \address{$^{(5)}$ Department of Applied Chemistry, University 
of Tokyo, Tokyo 113, Japan} %
 
\maketitle
%
%
\begin{abstract}

  The {\em ab}-plane optical spectra of two single crystals of underdoped 
La$_{2-x}$Sr$_{x}$CuO$_{4}$ were investigated. The reflectivity of  
La$_{1.87}$Sr$_{0.13}$CuO$_{4}$  has been measured in the frequency range 30 
-- 9,000~cm$^{-1}$ (0.004 -- 1~eV) both parallel  and 
perpendicular to the CuO$_2$ planes, whereas La$_{1.86}$Sr$_{0.14}$CuO$_{4}$  
was studied only in the $ab$-plane. The extended Drude model shows that the 
frequency-dependent effective scattering rate $1/\tau(\omega,T)$ is strongly 
suppressed below the high-frequency straight-line extrapolation, a signature 
of the  pseudogap state. This suppression can be seen from  temperatures 
below the superconducting transition up to 400~K. In the case of underdoped 
LSCO  the straight-line extrapolation is temperature independent below 
200~K, whereas above 200~K there is a strong temperature dependence of the 
high-frequency $1/\tau(\omega,T)$. The out-of-plane direction is also 
examined  for evidence  of the pseudogap state. 

\end{abstract} 
]

The presence of a pseudogap in the normal state of underdoped high 
temperature superconductors is by now widely accepted.\cite{timusk98} The 
strongest evidence for the pseudogap state comes from recent measurements of 
angle resolved photoemission spectra\cite{arpes} as well as vacuum 
tunneling\cite{tunneling}. However, these techniques both demand extremely 
high surface quality and have therefore mainly been restricted to Bi$_2$Sr$_2$CaCu$_2$O$_{8+\delta}$(Bi221) and 
YBa$_2$Cu$_3$O$_{7-\delta}$(Y123) materials, both with two CuO$_2$ layers per unit cell. Techniques that 
probe deeper into the sample such as dc transport\cite{bucher93,ito93,batlogg}, 
optical conductivity \cite{basov96,puchkov96a,puchkov96b} and NMR\cite{millis} 
were not only the earliest to show evidence of the pseudogap but have been 
extended to a much larger variety of materials, including several materials with one CuO$_2$.\cite{batlogg}  In all cases evidence for a pseudogap has been 
reported. 

The pseudogap in LSCO as seen by  NMR and neutron 
scattering\cite{mason} is rather weak and  has led to the suggestion that  
the existence of the pseudogap in the spin excitation spectrum is only 
possible in bilayer compounds such as Y123 and YBa$_2$Cu$_4$O$_4$(Y124). In particular, Millis 
and Monien attribute the pseudogap  (or the spin gap) to strong 
antiferromagnetic correlations between the planes in the bilayer, which are 
responsible for a quantum order-disorder transition.\cite{millis} 

Apart from having only one CuO$_2$ layer La$_{2-x}$Sr$_{x}$CuO$_4$ (LSCO)  
is also a good model system for the study of doping dependences since it can be 
doped by the addition of strontium over a wide range: from the underdoped,  
where $T_c$ increases with Sr content, to the optimally doped where $T_c$ 
reaches its maximum value of $\approx 40$ K at $x=0.17$, and  to the 
overdoped region where $T_c\rightarrow 0 $ at $x=0.34$.\cite{batlogg} 

The characteristic signatures of the pseudogap state in the dc 
resistivity\cite{bucher93} are seen clearly in  LSCO\cite{batlogg,uchida}. 
These are the striking deviations below a temperature $T^{\ast}$ from the 
high temperature linear resistivity, resulting in a clear break in slope at 
$T^*$. It was found by  B.~Batlogg {\it et al.} \cite{batlogg} that in LSCO 
$T^{\ast}$ decreases from 800~K to approximately 300~K as the doping level 
is increased from the strongly underdoped to just over the optimal doping 
level. Similar behavior at $T=T^{\ast}$ has been observed in the Hall effect 
coefficient and the magnetic susceptibility.\cite{Hwang,hall effect} 

\begin{figure}[t]
\leavevmode
\epsfxsize=\columnwidth
\centerline{\epsffile{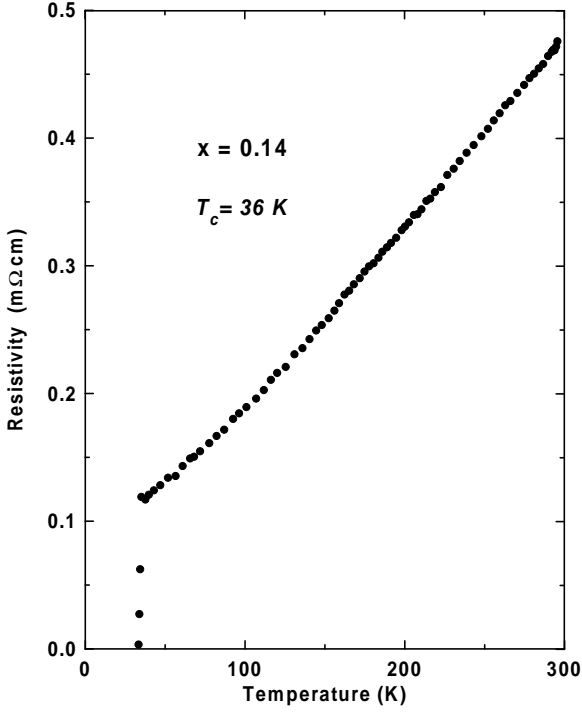}}
\caption{The temperature dependence of the in-plane resistivity of 
La$_{1.86}$Sr$_{0.14}$CuO$_4$ is shown with a sharp superconducting 
transition at ~36~K. The shape of the curve is consistent with 
$T^{\ast}$ being greater than 300~K.}\label{resistivity} \end{figure} 

The pseudogap can also be observed if the conductivity is measured in the 
frequency domain,  $\sigma(\omega)$,  where it shows up as a striking  
depression of the frequency dependent scattering rate at low frequency. 
It is found that below a frequency $\Omega_p \approx 600$ cm$^{-1}$, the 
scattering rate drops below its high temperature, high frequency, linear 
behavior. This effect has been clearly identified in the bilayer 
materials.\cite{basov96,puchkov96a,puchkov96b} One of the aims of this paper 
is to see if this behavior can also be observed in LSCO. A pseudogap state 
can be defined in terms of this suppression of scattering: the material is 
in the pseudogap state when the scattering rate falls below the high 
frequency straight-line extrapolation. In the low frequency limit the 
scattering rate is proportional to the dc resistivity. Due to this, the  
1/$\tau$($\omega$,T) suppression should be compared to the suppression of 
$\rho_{DC}$(T) \cite{batlogg} at   temperatures below the linear T 
dependence region. The IR measurement gives us the possibility to see both 
the frequency and the temperature dependence of this feature.

\begin{figure}[tbp]
\leavevmode
\epsfxsize=\columnwidth
\centerline{\epsffile{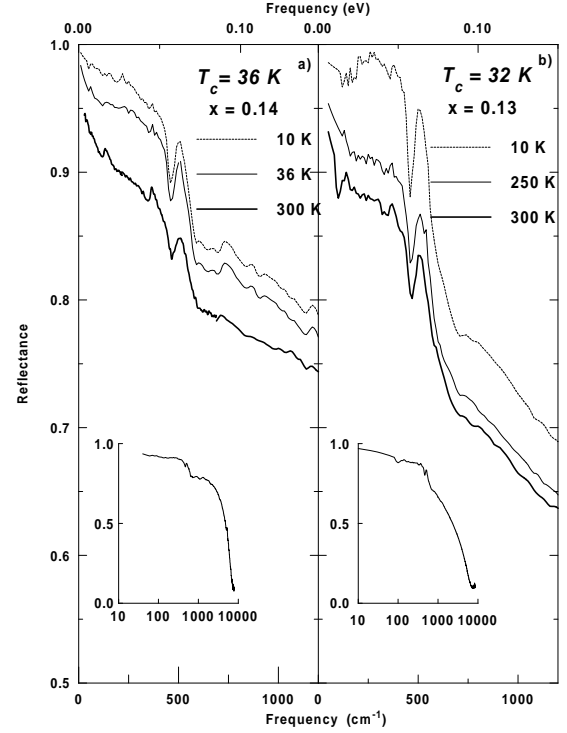}}
\caption{The reflectance of 
La$_{1.86}$Sr$_{0.14}$CuO$_4$  (a) and 
 La$_{1.87}$Sr$_{0.13}$CuO$_4$ (b) is shown. The solid lines show normal state spectra, the dashed curve shows superconducting state spectrum. The thinest line shows the spectrum at the temperature closest to T$_c$.
 The insert in  the left panel is a semi-log graph of the reflectance at 300~K which 
shows a  plasma edge around 7000~cm$^{-1}$.} \label{ab:reflectance}\end{figure}

A  pseudogap feature can also be observed in the {\it c-axis} IR {\it 
conductivity} in the form of a gap-like region of depressed 
conductivity at low frequency. It has been reported in 
YBa$_2$Cu$_3$O$_{7-x}$ (Y123) and YBa$_2$Cu$_4$O$_8$ (Y124) 
materials\cite{basov96,homes} as well as in LSCO\cite{c-axis 
LSCO,uchida c-axis}. In slightly underdoped LSCO the pseudogap 
state in the $c$-axis direction  is not as well 
defined as it is in the two plane materials.\cite{basov96} However, as the 
doping is reduced further, the $c$-axis pseudogap state features  below 0.1 eV 
become clearer.\cite{uchida c-axis} 

\begin{figure}[tbp]
\leavevmode
\epsfxsize=\columnwidth
\centerline{\epsffile{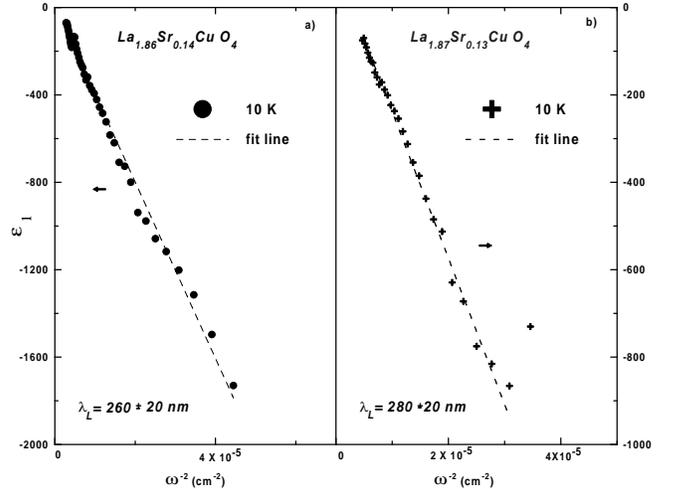}}
\caption{The real part of the dielectric function as a function of 
$\omega^{-2}$ of La$_{1.86}$Sr$_{0.14}$CuO$_4$  at 10~K  is shown in the panel 
a) and of La$_{1.87}$Sr$_{0.13}$CuO$_4$ at 25~K  is shown at the panel b). The 
dash line is linear fit. The slope of the fit gives the values of  
 the London penetration depth. }\label{epsilon}
\end{figure}

\begin{figure}[tbp]
\leavevmode
\epsfxsize=\columnwidth
\centerline{\epsffile{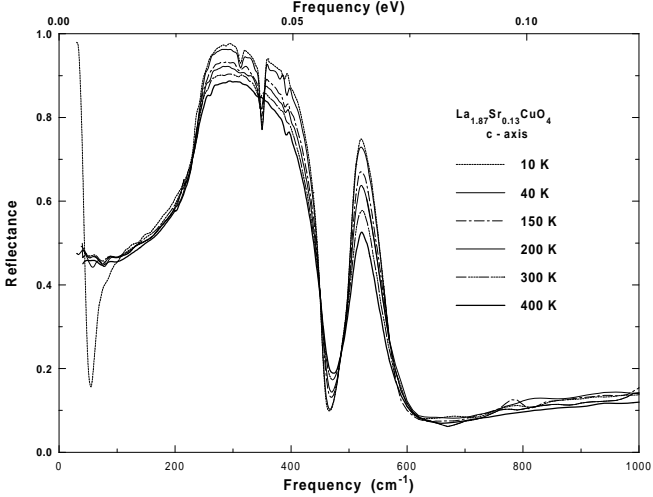}}
\caption{The reflectance of 
La$_{1.87}$Sr$_{0.13}$CuO$_4$ with $E \parallel c$ axis is shown. The temperature sequences are 10~K, 
40~K, 150~K, 200~K, 300~K and 400~K} \label{c:reflectance}\end{figure}

Previous work on the in-plane $\sigma(\omega)$ of the single layer 
lanthanum strontium cuprate includes work on the oxygen doped La$_2$CuO$_{4-
\delta}$\cite{Quijada}, thin films of LSCO\cite{Gao} as well as work done on LSCO single 
crystal at room temperature\cite{uchida}. To our knowledge, a study of 
the temperature  and doping dependence has not been done. We fill this gap 
here by performing optical measurements on high-quality LSCO single crystals 
at temperatures ranging from 10~K to 300~K at two different doping levels, 
both slightly underdoped.  Also the optical properties of both the ab-plane and 
c-axis of La$_{1.87}$Sr$_{0.13}$CuO$_4$ were measured on the same crystal. 

To better display the effect of increased coherence on $\sigma(\omega)_{ab}$ 
resulting from the formation of the pseudogap state we use the memory 
function, or extended Drude analysis.  In this treatment the complex 
optical conductivity is modeled by a Drude spectrum with a frequency-dependent scattering rate and an effective electron mass.\cite{goetze,allen} 
While the optical conductivity tends to emphasize free particle behavior, a 
study of the frequency dependence of the effective scattering rate puts more 
weight on displaying the interactions of the free particles with the 
elementary excitations of the system.\cite{gold} The temperature evolution 
of the frequency dependent scattering rate and effective mass spectra are of 
particular interest and are  defined as follows: 

\begin{equation} 
1/\tau(\omega, T) = {\omega^{2}_p\over4\pi} Re({1\over\sigma(\omega, T)})
\end{equation}
\begin{equation} 
{m^{\ast}(\omega, T)\over m_e} = {1\over \omega}{\omega^{2}_p\over4\pi} Im({1\over\sigma(\omega, T)}) 
\end{equation} 

Here, $\sigma(\omega, T)=\sigma_{1}(\omega ,T)+i\sigma_{2}(\omega, T)$ is 
the complex optical conductivity and $\omega_p$ is the plasma 
frequency of the charge carriers. 

The single crystals of La$_{2-x}$Sr$_x$CuO$_4$ with approximate dimensions 
5x3x3~mm$^3$ were grown by the traveling-solvent floating zone technique at 
Oak Ridge \cite{mook} in the case of $x = 0.14$ and in Tokyo \cite{kimura} 
in the case of $x = 0.13$.   The critical temperature was determined by both 
SQUID magnetization and resistivity measurements and was found to be 36~K 
for the nominal concentration of Sr $x = 0.14$ and 32~K for $x = 0.13$. 
Since the highest $T_c$ in the LSCO system has been found  to be 40~K for 
$x=0.17$, we conclude  that both  crystals are  underdoped.

\begin{figure}[tbp]
\leavevmode
\epsfxsize=\columnwidth
\centerline{\epsffile{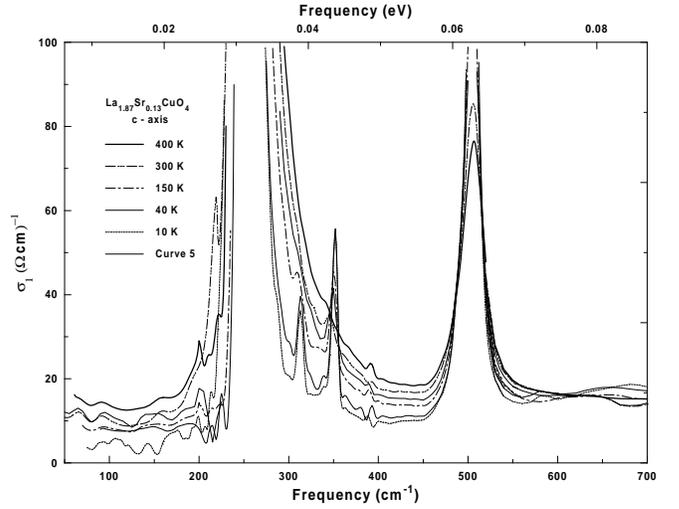}}
\caption{The c-axis conductivity of the La$_{1.87}$Sr$_{0.13}$CuO$_4$  is 
shown at different temperatures. Since the phonon peaks are dominant in the 
direction perpendicular to CuO$_2$ planes, the graph is focused at the 
background conductivity. Two inserts are  the c-axis conductivity of the underdoped La$_{1.87}$Sr$_{0.13}$CuO$_4$ measured  at 450~cm$^{-1}$ and  at 600~cm$^{-1}$. The c-axis conductivity at 450~cm$^{-1}$ is depressed below 300~K, however, it is somewhat constant above 600~cm$^{-1}$. This could be a signature of the pseudogap formation  at the temperatures less than 300~K and with the size of 500~cm$^{-1}$.} \label{c:cond}\end{figure}

\begin{figure}[tbp]
\leavevmode
\epsfxsize=\columnwidth
\centerline{\epsffile{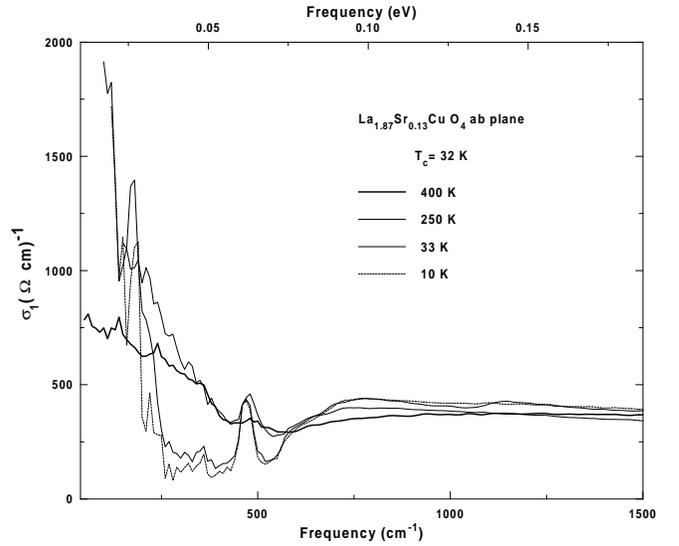}}
\caption{The ab-plane conductivity  of the La$_{1.87}$Sr$_{0.13}$CuO$_4$  is shown at different temperatures.}
\label{ab_cond13}
\end{figure}

The crystal with $x =0.14$ was aligned using Laue diffraction and polished 
parallel to the  CuO$_2$ planes. The crystal with $x = 0.13$ was polished 
in Tokyo to yield both ab-plane and ac-plane faces. Both surfaces were 
measured.  Polarizers were used for the ac-face data to separate the 
contribution of CuO$_2$ planes from the c-axis optical response. 

To get an uncontaminated ab-plane measurement, it is  important to have the 
sample surface accurately parallel to the {\it ab}-plane to  avoid any c-
axis contribution to the optical conductivity.\cite{miscut}  The miscut 
angle between the polished surface normal and the c-axis was checked by a 
high precision triple axis x-ray diffractometer  and was determined to be 
less than 0.8\%. 

All reflectivity measurements were performed with a Michelson interferometer 
using three different detectors which cover frequencies  ranging from 10 to  
10000 cm$^{-1}$ (1.25~$m$eV -- 1.25~eV). The experimental uncertainty in the 
reflectance data does not exceed 1$\%$. The dc resistivity measurements were 
carried out using a standard 4-probe technique. 

\begin{figure}[tbp]
\leavevmode
\epsfxsize=\columnwidth
\centerline{\epsffile{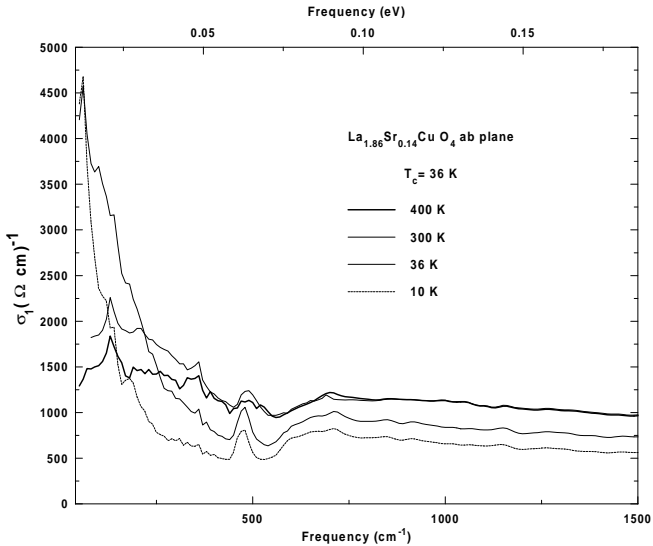}}
\caption{The ab-plane conductivity of La$_{1.86}$Sr $_{0.14}$CuO$_4$  
is shown at different temperatures.}
\label{ab_cond14}
\end{figure}

The result of the resistivity measurement on the same 
La$_{1.86}$Sr$_{0.14}$CuO$_4$  single crystal used in the optical 
measurements is shown in Fig.~\ref{resistivity}. It is commonly accepted that the DC-resistivity is linear at high temperatures for LSCO and that the pseudogap 
begins to form near the temperature where the resistivity drops below this 
linear trend.\cite{batlogg} At lower temperatures there is a region of 
superlinear temperature dependent resistivity. The $T^{\ast}$ value for our 
samples with $x=0.13$ and $x=0.14$ extracted from the phase diagram of 
Batlogg {\it et al.},\cite{batlogg} are 650~K and 450~K, respectively. In 
agreement with this, the   resistivity shows a superlinear  temperature 
dependence below  room temperature as expected in the pseudogap region. 

In Fig.~\ref{ab:reflectance}  we present the reflectivity data at temperatures above and 
below $T_c$ for the two samples. For clarity, only three temperatures are 
shown: $T=300$~K, an intermediate temperature  above the superconducting  transition 
and a low temperature $\approx 10$~K  in the superconducting state.  In the frequency region shown 
the reflectance  is strongly temperature dependent for both materials, 
dropping  by approximately  $10\%$ as temperature is increased from the 
lowest temperature to $T=300$~K. The plasma edge is observed at 
7800~cm$^{-1}$ (see insert of Fig.~\ref{ab:reflectance}). The distinct  peaks at approximately 
135 and 365~cm$^{-1}$  in the  LSCO  reflectivity spectra correspond to 
the excitation of ab-plane $TO$ phonons and the peak at 500~cm$^{-1}$ corresponds to a $LO$ phonon.\cite{tajima_phonons}    As in all 
other HTSC materials, the ab plane has a coherent response with very high reflectance.

\begin{figure}[h]
\leavevmode
\epsfxsize=\columnwidth
\centerline{\epsffile{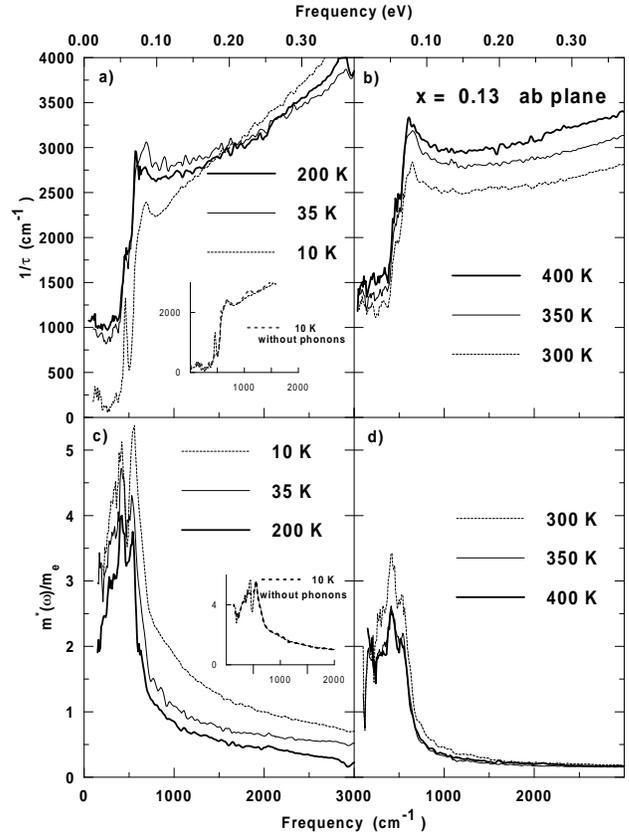}}
\caption{ Top panel: the low temperature  frequency dependent scattering rate of 
La$_{1.87}$Sr$_{0.13}$CuO$_4$ below T$^{\ast \ast}$(a) and above T$^{\ast \ast}$(b) is  calculated using Equation (1). The onset of the suppression 
in   a conductivity corresponds to a drastic change in the frequency 
dependence of the scattering rate below T$^{\ast}$. Above 700~cm$^{-1}$ the scattering 
rate is nearly temperature independent and has a linear frequency 
dependence below T$^{\ast \ast}$. 
Below 700~cm$^{-1}$ the scattering rate varies as $\omega^{1+\delta}$ 
and shows a strong temperature dependence. Bottom panel: The effective mass of La$_{1.87}$Sr$_{0.13}$CuO$_4$ below T$^{\ast \ast}$(a) and above T$^{\ast \ast}$(b)  is calculated using Equation 
(2). The onset of enhancement of ${m^{\ast}(\omega)\over m_e}$ 
corresponds to the onset of the suppression of the scattering rate. } 
\label{t13}
\end{figure}

The complex optical  conductivity $\sigma(\omega)$  was obtained by Kramers-Kronig analysis of the reflectivity data. Since, in principle, this analysis 
requires knowledge of the reflectance at all frequencies,  reflectivity 
extensions must be  used at high and low frequencies. The Hagen-Rubens 
formula was used for the low frequency reflectivity extrapolation, with 
parameters taken from the dc resistivity measurements on the same sample 
with $x=0.14$ shown in Fig.~\ref{resistivity} and the results  of H.~Takagi {\em et 
al.}\cite{takagi} for the sample with $x=0.13$. For the high-frequency 
extension for $\omega>8000$~cm$^{-1}$ we used the reflectivity results of 
Uchida {\it et al.}\cite{uchida} At  frequencies higher than 40~eV the 
reflectivity was assumed to fall as $1/\omega ^{4}$. 

We calculate the plasma frequency of the superconducting charge carriers and 
the London penetration depth using the following formula:\cite{Timusk 
review} \begin{equation} \epsilon_1 = 1 - {\omega_{ps}^{2}\over\omega^2}. 
\end{equation} 

The slope of the low-frequency  dielectric function, 
$\epsilon_1(\omega)$, plotted  as a function of $w^{-2}$ in Fig.~\ref{epsilon}a,b  
gives  plasma frequencies of 6100~cm$^{-1}$ and 5700~cm$^{-1}$ 
in the superconducting state. The corresponding London penetration 
depths are $\lambda_L=1/2\pi\omega_{ps}=250$~nm and 280~nm for 
La$_{1.86}$Sr$_{0.14}$CuO$_4$ and La$_{1.87}$Sr$_{0.13}$CuO$_4$, 
respectively. These values  are in  good agreement with those 
obtained previously by Gao {\it et al.}  in films\cite{Gao}   
($\lambda_L = 275\pm5~nm$)  and by muon-spin-relaxation\cite{msr}  
($\lambda_L = 250~nm$). 
\

The c-axis reflectance of the $x=0.13$ sample is shown in Fig.~\ref{c:reflectance}. The 
corresponding conductivity is low and is dominated by optical 
phonons (Fig.~\ref{c:cond}).

In YBCO 123 and 124 the pseudogap in c-axis conductivity manifests itself as 
a depression in conductivity at low frequency.\cite{homes,basov96,Homes95} 
There is no coherent Drude peak and  the conductivity is flat and frequency 
independent. In  the temperature and doping range where a pseudogap is 
expected a low frequency depression of conductivity is seen with an edge in 
the 300-400 cm$^{-1}$ region where the conductivity rises to the high 
frequency plateau. 

In order to isolate the electronic features of our LSCO c-axis spectrum we 
magnify the low value region of $\sigma_{1c}$ (Fig.~\ref{c:cond}).  There is no sharp 
pseudogap edge in the low-frequency infrared data for underdoped LSCO as there is in the case of Y123. It is possible that such a feature could be hidden under 
the large phonon structure. Efforts to subtract the phonons in order to 
extract the background conductivity were found to be extremely sensitive to 
the choice of their shape in fitting procedures. Nonetheless, the raw data 
clearly show that  there is low frequency depression of the c-axis 
conductivity. Conductivity at 450~cm$^{-1}$ is uniformly suppressed below 
T$= $300 K (Fig.~\ref{c:cond}insert), whereas the conductivity at 600 cm$^{-1}$ is nearly 
constant at all temperatures. Based on this analysis one can 
conclude that the pseudogap state in the c-axis opens up below 300~K and its 
size is approximately equal to 500~cm$^{-1}$. 

\begin{figure}[tbp]
\leavevmode
\epsfxsize=\columnwidth
\centerline{\epsffile{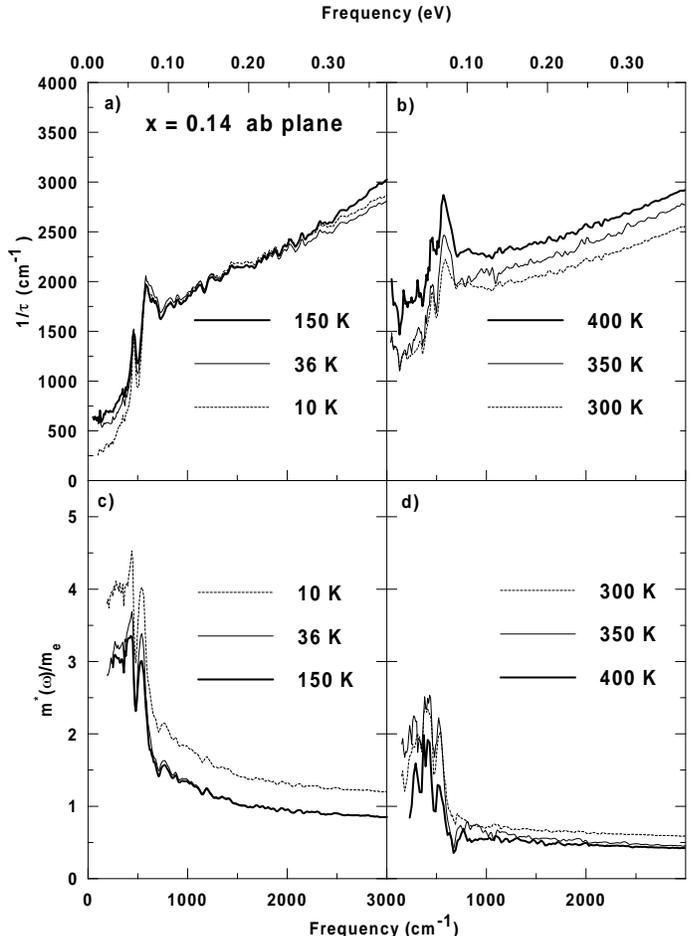}}
\caption{ Top panel: the high temperature effective scattering rate of 
 La$_{1.86}$Sr$_{0.14}$CuO$_4$ below T$^{\ast \ast}$ (a) and  above T$^{\ast \ast}$ 
(b) is  calculated using Equation (1).  Above 700~cm$^{-1}$ the scattering 
rate  has a linear frequency  and temperature independent below T$^{\ast \ast}$ (a) and temperature dependent above T$^{\ast \ast}$(b). 
Below 700~cm$^{-1}$ the scattering rate varies as $\omega^{1+\delta}$ 
and shows a strong temperature dependence. Bottom panel: The effective mass of La$_{1.87}$Sr$_{0.13}$CuO$_4$ below T$^{\ast \ast}$(a) and above T$^{\ast \ast}$(b) samples is calculated using Equation 
(2). } 
\label{t14}
\end{figure}

Manifestations of the pseudogap in the $ab$-plane conductivity exist 
as a loss of spectral weight between 700 and 200 cm$^{-1}$ 
balanced by increases both below and above this frequency. In both Fig.~\ref{ab_cond13} 
and Fig.~\ref{ab_cond14} one can see the temperature evolution of the sharp 
depression in ab-conductivity below 700 cm$^{-1}$ at 
temperatures above T$_{c}$. A much clearer picture of the 
pseudogap state can be seen from the effective scattering rate, 
1/$\tau$($\omega$,T), calculated from the conductivity using  
equation~(1)  which is shown along with the effective mass in 
Fig.~\ref{t13} and Fig.~\ref{t14}.  The  1/$\tau$($\omega$,T) spectra can 
conveniently be divided into two regions.  In the high frequency 
region, starting at about  700~cm$^{-1}$, the scattering rate 
varies linearly with frequency while in the low frequency region 
there is a clear suppression of  1/$\tau$($\omega$,T) below this 
linear trend.\cite{anton_review} The temperature where this 
suppression first appears serves as a definition of $T^{\ast}$, 
the onset temperature of the pseudogap state. As the temperature 
is lowered below $T^{\ast}$ this suppression becomes deeper.   
We find that for underdoped La$_{2-x}$Sr$_{x}$CuO$_4$  $T^{\ast} 
\geq$ 400~K, an order of magnitude higher than the 
superconducting transition temperature $T_c$  (32~K). This  is 
significantly different from previous results on  cuprates where 
T$^{\ast}$ more or less coincides with T$_c$ near optimal 
doping. 

The temperature dependence above 700~cm$^{-1}$ is strongly 
influenced by the level of Sr doping. In the underdoped sample 
the high frequency scattering rate is nearly temperature 
independent up to a certain temperature (Fig.~\ref{t13}a and Fig.~\ref{t14}a), 
which we will call $T^{\ast\ast}$ above which a pronounced 
temperature dependence of 1/$\tau$($\omega$,T) is seen (Fig.~\ref{t13}b 
and Fig.~\ref{t14}b). In the $x=0.13$  sample 
$T^{\ast\ast}\approx 200$ K while in the $x=0.14$ sample 
$T^{\ast\ast}\approx 150$ K. In the overdoped samples  the 
scattering rate above 700 cm$^{-1}$ increases uniformly with 
temperature\cite{startseva2} at all temperatures suggesting 
$T^{\ast\ast}\rightarrow 0$ in that limit.  This behavior is 
also seen in  other overdoped HTSC.\cite{anton_review} 
 
If one extrapolates the 300 K scattering rates to zero frequency 
one finds that for the $x=0.13$ sample the scattering rate $1/\tau_0\approx 
2500$ cm$^{-1}$ and for the $x=0.13$ sample this rate is 
$\approx 1500$~cm$^{-1}$. These scattering rates are much higher 
than what is seen for the higher $T_c$ materials reviewed by 
Puchkov {\it et al.}\cite{anton_review} where at 300 K $1/\tau_0\approx 
1000 \pm 200$ cm$^{-1}$ for several families and many doping 
levels. This high residual scattering differentiates the LSCO 
material from the other cuprates.

If we call the frequency below which the scattering rate is 
suppressed the $ab$-plane pseudogap $\omega_{ab} \approx 700$ 
cm$^{-1}$ we find it is clearly bigger than  the c-axis pseudogap 
frequency $\omega_{c} \approx 500$ cm$^{-1}$

\begin{figure}[tbp]
\leavevmode
\epsfxsize=\columnwidth
\centerline{\epsffile{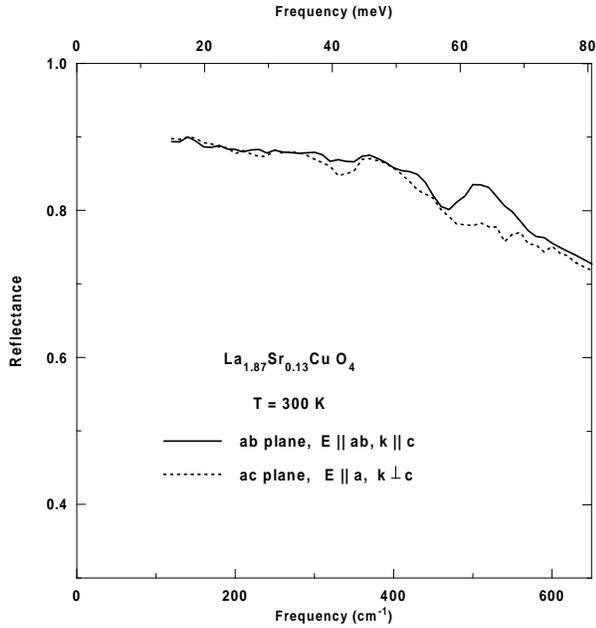}}
\caption{ Comparison of the reflectance of La$_{1.87}$Sr$_{0.13}$CuO$_4$  measured from ab plane with $k \parallel c$ and from ac plane with $ k \perp c$.} 
\label{maureen:ref}
\end{figure}

\begin{figure}[tbp]
\leavevmode
\epsfxsize=\columnwidth
\centerline{\epsffile{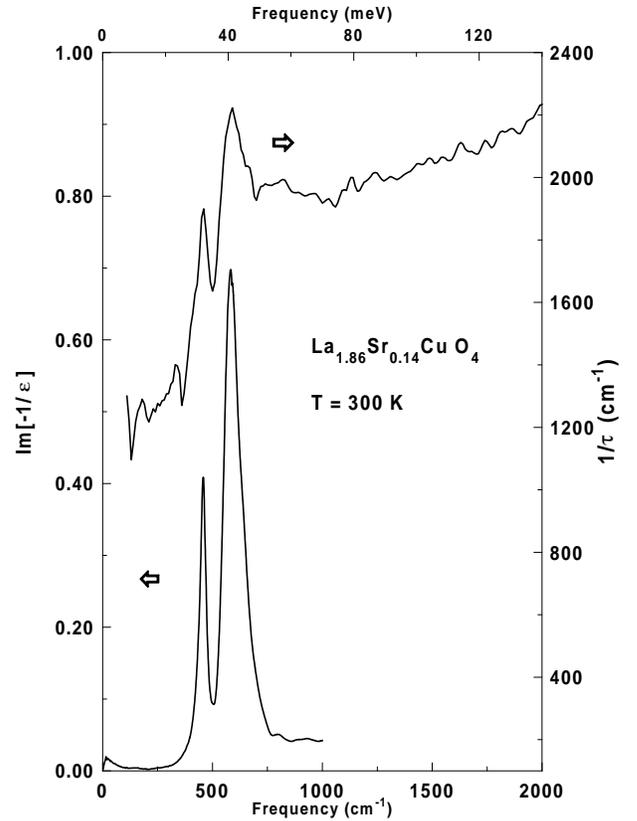}}
\caption{ Comparison of the peaks position at 450 and 580~cm$^{-1}$ in the dielectric loss function of the c-axis phonons with the  effective scattering rate of 
 La$_{1.86}$Sr$_{0.14}$CuO$_4$ at room temperature. The correspondence of the peaks positions, width and the relative strength suggests that the nature of the peak may lie in the coupling of a ab-plane spectra to the c-axis longitudinal optical phonons.} 
\label{maureen:t}
\end{figure}

In addition to the pseudogap depth and temperature dependence,  
several other features of Figures 8 and 9  should be mentioned. 
The position of the pseudogap remains at 700~cm$^{-1}$ for all 
temperatures. There are also several peaks positioned at 500~cm$^{-1}$ 
in the scattering rate which complicate the analysis, 
particularly in the case of the  sample with $x=0.14$. These 
peaks have been observed by other groups and have been 
attributed to polaronic effects.\cite{bipolarons,Thomas92}  
Another possible explanation is the correlation of the ab-plane 
conductivity with c-axis LO 
phonons.   We did observe the difference in 
 contribution of LO phonons to the ab plane reflectance with 
different propagation directions, an effect first observed by 
Reedyk {\it et al.},\cite{reedyk92} and 
also seen in the $k\parallel c$ vs. $k\perp c$ spectra obtained by 
Tanner's group.\cite{Quijada} In Fig.~\ref{maureen:ref} the reflectance with $E 
\parallel a$ and $k\parallel c$, is compared with the 
reflectance with  
$E \parallel ab$ and $k\parallel a$, with the 
La$_{1.87}$Sr$_{0.13}$CuO$_4$ sample at room temperature. There is 
an extra feature observed at 500~cm$^{-1}$ in the spectra with 
$k\parallel c$. Further evidence that the c-axis LO phonons 
couples to ab plane features can be seen in Fig.~\ref{maureen:t}. The 
comparison between peaks in the effective scattering rate at 
450~cm$^{-1}$ and 580~cm$^{-1}$ to the peaks in Im(-
1/$\epsilon_{c}$) shows the same strong correlation seen in many 
other cuprates.\cite{reedyk92}  

For completeness we also plot the effective mass of underdoped 
samples at  low temperatures(Fig.~\ref{t13}c) and  high 
temperatures(Fig.~\ref{t14}c,d). As expected,  ${m^{\ast}(\omega)\over 
m_e}$ rises to a maximum $\approx 3$ forming a peak at $\approx 
400$ cm$^{-1}$. The enhancement of the effective mass in the 
pseudogap state  as well as  the upper limit of 
${m^{\ast}(\omega)\over m_e}$ are similar to what has been 
previously reported for  Y123, Y124 and 
Bi2212.\cite{anton_review} 

Before closing we compare our results with data of Gao {\it et 
al.}\cite{Gao} on La$_{2-x}$Sr$_{x}$CuO$_{4+\delta}$ films and Quijada {\it 
et al.}\cite{Quijada} on  oxygen doped La$_2$CuO$_{4+\delta}$. Our results 
in the underdoped case are comparable with those of the oxygen doped 
material, although Quijada {\it et al.} did not carry out a frequency 
dependent scattering rate analysis for their underdoped sample.  The film 
results of Gao {\it et al.} are  quite different from our findings.  The 
films used in that study had a strontium level that would correspond to 
optimal doping  in crystals. However, the $1/\tau(\omega)$ curves deviate 
markedly from what we observe for  slightly under and overdoped samples. The 
authors performed an extended Drude analysis and  found a strongly 
temperature dependent scattering rate even at low temperatures. This is  in 
sharp contrast to our results which would suggest a very weak temperature 
dependence in this temperature region. Based on our work, their samples 
should be in the pseudogap state since they have an $x$ value near optimal 
doping. Comparing these results with other systems, in particular with 
Tl2202, two factors suggest the possibility that the films may be overdoped. 
First, their $T_c$ was near 30 K, lower than that expected for optimal 
doping. Secondly, it is known that the oxygen level in films can vary 
substantially and in LSCO oxygen can have a major influence on the doping 
level\cite{zhang}. On the other hand, we cannot completely rule out the 
possibility that all the crystal results are affected by the polishing 
process, and that the films better represent  the bulk material. It is 
clearly important to measure films where the oxygen content is controlled by 
selective annealing. 

In  conclusion, the optical data in the  far-infrared region, taken on two  
underdoped {\it single-layered} high-$T_c$ superconductors, shows  clear 
evidence of {\it a pseudogap state} in both the scattering rate and 
conductivity along CuO$_2$ planes. This pseudogap state  extends to  higher 
temperatures than that observed in the multi-layered underdoped cuprates 
such as YBCO and BSCO. 

The scattering rate  is similar for both systems in the pseudogap state. At 
low frequencies, $\omega \leq$ 700~cm$ ^{-1}$, the scattering rates are {\em 
temperature} dependent and change with  frequency in a {\em non-linear} 
fashion. Above 700 cm$^{-1}$ this behavior becomes  {\em linear}. Within 
experimental uncertainty the observed high frequency scattering rate of the 
underdoped sample is {\em not affected by temperature} up to certain 
temperature T$^{\ast \ast}$. This temperature is equal to 200~K in case of  
La$_{1.87}$Sr$_{0.13}$CuO$_{4}$ and 150~K in case of 
La$_{1.86}$Sr$_{0.14}$CuO$_{4}$ . Above  T$^{\ast \ast}$ the high frequency 
scattering rate is temperature dependent. This behavior is identical to 
the high-frequency effective scattering rate of an overdoped 
HTSC.\cite{anton_review}  

Our findings in the direction perpendicular to  the CuO$_2$ 
planes showed that the depression of the c-axis conductivity is not as 
prominent as the one found in the two layer HTSC. Nevertheless, the 
signature of the pseudogap can be seen at the frequencies below 500~cm$^{-
1}$ up to room temperature.

We would like to thank J.D.~Garrett for  help in  aligning the sample 
and also P.C.~Mason, M.~Lumsden and B.D.~Gaulin for  determining  the 
miscut angle of the underdoped LSCO crystal.  We also take this 
opportunity to thank K.C.~Irwin and J.G.~Naeini for their useful 
collaboration. This work was supported by the Natural Sciences and 
Engineering Research  Council of Canada and The Canadian Institute 
for Advanced Research.


%

\end{document}